\documentclass[nofootinbib,aps,prd,twocolumn,showpacs,preprintnumbers,amsmath,amssymb]{revtex4}
\usepackage{epsfig}

\newcommand{\tr}{{\rm Tr}}
\newcommand{\non}{\nonumber}
\def\lsim{\raise0.3ex\hbox{$<$\kern-0.75em\raise-1.1ex\hbox{$\sim$}}}
\def\gsim{\raise0.3ex\hbox{$>$\kern-0.75em\raise-1.1ex\hbox{$\sim$}}}

\newcommand{\nc}[1]{\newcommand{#1}}
\nc{\bo}[1]{\mbox{\boldmath \( #1 \! \! \)  \unboldmath}}
\nc{\be}{\begin{eqnarray}}
\nc{\ee}{\end{eqnarray}}
\nc{\bew}{\begin{eqnarray*}}
\nc{\eew}{\end{eqnarray*}}
\nc{\bs}{\begin{subeqnarray}}   
\nc{\es}{\end{subeqnarray}}     
\nc{\nnn}{\nonumber \\}
\nc{\f}[2]{\frac{#1}{#2}}
\nc{\td}[2]{\f{d #1}{d #2}}
\nc{\pd}[2]{\f{\partial #1}{\partial #2}}
\nc{\suli}{\sum\limits}
\nc{\proli}{\prod\limits}
\nc{\ili}{\int\limits}
\nc{\sr}[2]{\stackrel{#1}{#2}}
\nc{\dps}{\displaystyle}
\nc{\ket}[1]{\left| #1 \right>}
\nc{\bra}[1]{\left< #1 \right|}
\nc{\bracket}[2]{\left< #1 \right| \left. \! #2 \right>}
\nc{\norm}[1]{\left\| #1 \right\|}

\begin{document}

\title{Color Screening and Quark-Quark Interactions in Finite Temperature QCD
}

\author{Matthias D\"oring$^\dagger$}
\author{Kay H\"{u}bner$^{\dagger\dagger}$}
\author{Olaf Kaczmarek$^\dagger$}
\author{Frithjof Karsch$^{\dagger\dagger}$} 
\affiliation{ 
$^\dagger$ Fakult\"{a}t f\"{u}r Physik, Universit\"{a}t 
Bielefeld, D-33615 Bielefeld, Germany\\
$^{\dagger\dagger}$ Physics Department, Brookhaven National Laboratory, 
Upton, New York 11973, USA}

\date{\today}
\preprint{BI-TP 2007/01}
\preprint{BNL-NT-07/5}


\begin{abstract}
We analyze the screening of static diquark sources in 2-flavor QCD and
compare results with the screening of static quark-antiquark pairs.
We show that a two quark system in a fixed color representation is screened
at short distances like a single quark source in the same color 
representation whereas at large distances the two quarks are screened
independently. 
At high temperatures we observe that the relative strength of the 
interaction in diquark and quark-antiquark systems, respectively,  
obeys Casimir scaling. We use this result to examine the 
possible existence of heavy quark-quark bound states in the high 
temperature phase of QCD. We find support for the existence of $bb$
states up to about $2T_c$ while $cc$ states are unlikely to be formed
above $T_c$.
\end{abstract}

\maketitle

\section{Introduction}\label{secintroduction}

Correlation functions of static quark and antiquark sources,
represented by the Polyakov loop and its hermitean conjugate, have 
proven to be very useful observables that allow to characterize 
thermal modifications of the interaction among partons at high temperature.  
At short distances Polyakov loop correlation functions are sensitive to the 
interaction
between quarks and antiquarks which is transmitted through gluon exchange. 
The thermal modification of gluons propagating in a hot medium
is reflected in the screening of the interaction among static quark
and antiquark sources. This allows, for instance, the study of the 
running of the QCD coupling \cite{Necco} and its temperature dependence
\cite{Zantow, Zan2}.

The large distance behavior of static quark correlation functions is 
sensitive to the confining properties of the QCD vacuum, the breaking
of the string due to pair creation in the vacuum as well as the
transition to a deconfined phase at high temperature and density.
While at low temperature the correlation functions calculated in a 
$SU(3)$ gauge theory approach zero at large distances and thus reflect 
strict confinement of quarks, they approach constant
values in QCD with light dynamical quarks indicating the onset of
string breaking at some characteristic distance.
At zero temperature string breaking arises through the
creation of a quark-antiquark pair from the QCD vacuum. With increasing
temperature more complex mechanisms involving quarks and gluons from the
thermal medium also start playing a role. In the high temperature phase
static quark correlation functions approach constant values in QCD as well
as in pure gauge theories suggesting that screening of static sources through
a gluon cloud is the dominant mechanism for the neutralization of the 
color charge of static external sources.

Information deduced from the temperature dependence of the change of free
energy induced in a thermal medium due to the presence of external sources
has been used to model the thermal modification of the potential between
heavy quarks. Through an analysis of the non-relativistic Schr\"odinger
equation, which may be a good approximation for heavy quarks, it has
been concluded that some heavy quark bound states may also survive the
transition to the high temperature phase of QCD 
\cite{Alberico:2006vw,Digal:2001iu,Digal:2001ue,Wong:2001uu,
Shuryak,Blaschke:2005jg,Alberico:2005xw,Mocsy:2005qw,Wong:2004zr}.
Recently it has been
argued that also more exotic, colored bound states made up of quark
antiquark pairs or quark-quark pairs may exist in a thermal medium 
at high temperature \cite{Shuryak}. To quantify such a scenario for 
quark-quark systems requires a better understanding of the interaction 
among quarks at high temperature and the screening of colored diquark 
states. 

First studies of quark-quark interactions at finite temperature
have been performed in quenched QCD \cite{Nakamura} and some results
for QCD with light staggered fermions have been reported at the Lattice
conference in 2005 \cite{Fodor,Hubner}. 
In this paper we will focus on an analysis of static quark-quark correlation
functions in 2-flavor QCD. We will calculate free energies of diquark
systems at finite temperature and compare with results obtained for
quark-antiquark systems at the same temperature. We will also calculate
the induced quark number that arises from the presence of a diquark
source that has non-vanishing triality in a thermal bath of quarks
and gluons.  This provides information on the response of the
thermal medium to the presence of external sources and allows to 
distinguish between the quark dominated screening mechanisms at low 
temperature (string breaking) and the gluon dominated screening at
high temperature. 

This paper is organized as follows. In the next Section we 
introduce the framework for our study of diquark correlation functions,
discuss the basic setup for our numerical calculations and present
some basic results on thermal properties of diquark and quark-antiquark
free energies.
In Section III we present results on free energies of color anti-triplet
and sextet diquark configurations calculated in Coulomb gauge and discuss 
the gradual change from  string breaking at low temperature to screening 
at high temperature which becomes transparent in the
behavior of the net quark number induced in a thermal medium due to
the presence of static quark sources. 
In Section IV we discuss consequences of the observed Casimir scaling
in diquark and quark-antiquark systems for the possible existence of
colored quark-quark bound states at high temperature. We finally
give our conclusions in Section V.

\section{Diquark free energies}\label{secsimdetails}
The general framework for the analysis of thermal properties of QCD in the 
presence of $n$ static quarks and $m$ static antiquarks has been formulated
by McLerran and Svetitsky \cite{McL}. 
We concentrate here on 2 quark systems for which
the lattice regularized QCD partition function with $n_f$ light dynamical 
quark degrees of freedom and static sources represented by the operator
$L_{QQ}^{(c)}(r)$ is given by,
\begin{eqnarray}
Z_{QQ}^{(c)}(\beta,\hat{m},r)\hspace{-0.1cm} &=&\hspace{-0.1cm}
\int \prod_{x,\nu} {\rm d}U_{x,\nu} L_{QQ}^{(c)}(r)
\left( {\rm det}\; D(\hat{m}, \mu)\right)^{n_f/4} \times \nonumber \\
~&&\hspace{1.0cm}{\rm e}^{-\beta S_G (U)} \;\; .
\label{QQ_partition}
\end{eqnarray}
where $U_{x,\nu} \in SU(3)$ are gauge field variables defined on 
the links of a 4-dimensional lattice of size $N_\sigma^3N_\tau$;  
$\beta = 6/g^2$ denotes the gauge coupling, $\hat{m}$ is the bare
quark mass for the $n_f$ degenerate quark flavors. Furthermore, 
$D(\hat{m},\mu)$ 
denotes the fermion matrix, for which we use the staggered fermion
discretization scheme. Although all our calculations have been performed
at vanishing quark chemical potential ($\mu$), we have indicated in 
Eq.~\ref{QQ_partition} the dependence of the fermion determinant on $\mu$. 
We will make use of this in our discussion of
the quark number induced in a thermal medium due to the presence of
external sources.

In Eq.~\ref{QQ_partition} $L_{QQ}^{(c)}(r)$ denotes the operator for two 
static quark sources separated by a distance $r$. The superscript $(c)$ 
indicates the color representation of the diquark system which we will
specify later. A static quark source is represented by the Polyakov loop, 
\begin{eqnarray}
P(\vec x) = \left( z_{n_f}(\beta,\hat{m})\right)^{N_\tau} 
\prod_{x_0=1}^{N_\tau} 
U_{(x_0,\vec x),0}\; ,
\label{polyakov}
\end{eqnarray}
and the source for antiquarks is given by the hermitean conjugate,
$P^{\dagger} (\vec x)$. In the definition of the Polyakov loop we have
introduced a renormalization constant,  $z_{n_f}(\beta,\hat{m})$, that removes 
divergent self energy contributions to the static quark sources and
insures that the diquark partition function normalized with the partition
function in the absence of any sources,
\begin{equation}
Z(\beta,\hat{m}) =
\int \prod_{x,\nu} {\rm d}U_{x,\nu} 
\left( {\rm det}\; D(\hat{m}, \mu)\right)^{n_f/4}
{\rm e}^{-\beta S_G (U)}  ,
\label{partition}
\end{equation}
has a well defined continuum limit. The
renormalization constants for static quark sources
have been determined previously for $n_f=0$ (quenched QCD) \cite{kac}
and 2-flavor QCD ($n_f=2$) \cite{Zan2} 
for the action and for values of the gauge coupling used 
also in this study.

The normalized $QQ$ partition function,
\begin{eqnarray}
C^{(c)}_{QQ}(r,T) & = & 
\frac{Z_{QQ}^{(c)}(\beta,\hat{m},r)}{Z(\beta,\hat{m})} 
=\left\langle L_{QQ}^{(c)}(r) \right\rangle \; ,
\label{eq:corr_fct}
\end{eqnarray}
is the correlation function for two static
quark sources separated by a distance $r$. Its logarithm defines
the free energy of the static diquark system, {\it i.e.} the change
in free energy of a thermal medium at temperature $T$ that arises form 
the presence of two static sources, 
\begin{eqnarray}
F^{(c)}_{QQ}(r,T)=-T\; \ln C^{(c)}_{QQ}(r,T) \;  .
\label{eq:fx}
\end{eqnarray}

We note that
the discussion of diquark partition functions and free energies given so 
far is completely analogous to the discussion usually given for the 
expectation value of the Polyakov loop\footnote{Here and everywhere else we
use the normalization of traces in color space $\tr$\boldmath$1$\unboldmath$= 1$.},
\begin{equation}
L_{Q} =  \frac{1}{N_\sigma^3}\sum_{\vec{x}}
\tr P(\vec{x})  \; ,
\end{equation}
and the related static quark free energy,
\begin{equation}
F_{Q}(T)=-T\; \ln  
\left\langle  
L_{Q}
\right\rangle
\; .
\label{eq:fQ}
\end{equation}
The Polyakov loop operator $L_{Q}$ simply has been replaced
by the diquark operator $L_{QQ}^{(c)}(r)$. 
 
Similarly one may, of course, also discuss the partition function and free
energies of quark-antiquark systems. 
To be specific we
introduce here the so-called color averaged operators\footnote{Note
that we always use renormalized operators for static quark sources.}
for a diquark and quark-antiquark system,
\begin{eqnarray}
L_{QQ}^{({\rm av})}(r) &=& \tr P(0)\tr P(r)\; ,\\
L_{\bar{Q}Q}^{({\rm av})}(r) &=& \tr P^\dagger(0)\tr P(r)\; .
\label{QQ_av}
\end{eqnarray}

Like the Polyakov loop also the
operator defining the diquark partition function is not invariant
under global $Z(3)$ transformations. The static quark and diquark
sources represent states of non-vanishing triality. 
Due to the $Z(3)$ symmetry of the $SU(3)$ gauge action partition
functions with non-vanishing triality will be zero in quenched 
QCD. To insure a proper definition of diquark partition functions
also for the pure $SU(3)$ gauge 
theory one should introduce a symmetry breaking term which is removed
again after taking the thermodynamic limit. 
In the presence of dynamical quarks the $Z(3)$ symmetry is, however,
explicitly broken and the partition functions with non-zero triality
are well defined. The fermion sector of the 
QCD partition function, Eq.~\ref{partition}, provides the appropriate 
number of quark and antiquark contributions such that
only states with vanishing triality will finally contribute to the
partition function. We will elaborate further on this in 
Section \ref{sec_string_breaking}.

\begin{figure}[t]
\epsfig{file=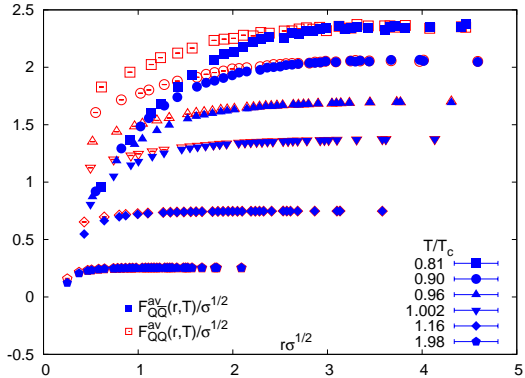,angle=270,width=7.5cm}
\caption{Diquark (open symbols) and quark-antiquark (filled symbols) 
free energy for color averaged sources separated at a distance expressed in 
units of the square root string tension $r\sqrt{\sigma}$ at temperatures 
below and above the transition temperature. 
}
\label{fig:plcqqav}
\end{figure}

The analysis of diquark 
free energies presented in this paper 
has been performed on data samples that have been used
previously to analyze free energies of static quark-antiquark systems
\cite{Zan2}. These data samples have been generated on $16^3\times4$
lattice for 2-flavor QCD using an improved staggered fermion action
and quark masses, that correspond to a light pseudo-scalar (pion) mass
of about $770$~MeV. For further details on this data sample and 
further details on the action and simulation parameters we refer to
\cite{Zan2} and references quoted therein.
The SU(3) pure gauge data used in Section \ref{subsec:sextet_screening} 
have been obtained on lattices of size $32^3\times 4$, where we used the
tree-level Symanzik improved gauge action \cite{Symanzik:1983gh}. For further details
see \cite{huebner}.  

In Fig.~\ref{fig:plcqqav} we compare the color averaged free energy of a 
static diquark system with that of a static quark-antiquark system both 
calculated in 2-flavor QCD. The figure shows free energies as 
function of the separation of the two sources measured in units of the
zero temperature string tension, $r\sqrt{\sigma}$. In both
cases renormalized Polyakov loop operators, Eq.~\ref{polyakov}, have been 
used; no further normalization of the relative magnitude of
$QQ$ and $\bar{Q}Q$ free energies thus is needed nor has it been performed. 
Apparently the free energies of a diquark and a quark-antiquark 
system approach the same large distance limit at all temperatures. 
In fact, as noted previously \cite{Zantow,Zan2},
this asymptotic value equals twice the free energy of a static
quark represented by the renormalized Polyakov loop, Eq.~\ref{eq:fQ}.
This observation confirms our picture of screening at large distance;
the two static sources are screened independently and the thermal bath
does not distinguish between quark and antiquark sources.

At short distances
the $\bar{Q}Q$ free energies drop faster than the $QQ$ free energies.
This is expected from a  perturbative analysis of the interaction in 2 
quark systems; the difference in Casimir factors for one gluon-exchange
in quark-quark and quark-antiquark systems
suggests that the attractive interaction in an anti-triplet  $QQ$ system is 
half as strong
as in a singlet $\bar{Q}Q$ system. The color averaged free energies
receive also contributions from octet and sextet representations of a 
$\bar{Q}Q$ and $QQ$ system, respectively. These contributions are, however,
repulsive and thus are exponentially suppressed in Polyakov loop correlation 
functions at short distances. An additional difference in free energies
of $QQ$ and $\bar{Q}Q$ arises from the fact that due to the non-vanishing 
triality of a $QQ$ state the free energy of a diquark system will receive an 
additional screening contribution which is not present in a $\bar{Q}Q$ 
system. 

To quantify the differences seen in Fig.~\ref{fig:plcqqav} and to explore 
in more detail the mechanism that leads to screening of color charges 
in a hot medium it is advantageous to study diquark and quark-antiquark 
correlation functions in definite color channels which can be defined in 
fixed gauges.

\section{Anti-triplet and sextet diquark free energies}

In addition to the color averaged diquark source introduced in Eq.~\ref{QQ_av}  
we will consider here sources in different color channels. This 
is analogous to studies of color singlet and octet free energies performed
for static quark-antiquark systems.
A system of two quarks can be either in a color anti-triplet (antisymmetric) 
or color sextet (symmetric) state,
\begin{equation}
 3 \otimes 3 = \overline{3}\oplus 6 \; .
\end{equation}
The corresponding operators defining the anti-triplet
and sextet partition functions are given by
\cite{Nadkarni:1986cz,Nadkarni:1986as}
\begin{eqnarray}
  L^{(\overline{3})}_{QQ}(r) & = & \frac{3}{2}\tr P(0)\tr P(r) -
  \frac{1}{2}\tr P(0)P(r)\; , 
\label{eq:part_fct1}\\\non &&\\
\label{eq:part_fct2}
  L^{(6)}_{QQ}(r) & = & \frac{3}{4}\tr P(0)\tr P(r) +
  \frac{1}{4}\tr P(0)P(r) \; .
\end{eqnarray}
These operators are, however, gauge dependent and the corresponding
partition functions can only be studied in a fixed gauge. We will 
analyze them in Coulomb gauge. 
The partition function for the color averaged diquark system can be written 
as a superposition of partition functions for color anti-triplet and 
sextet systems,
\begin{equation}
Z_{QQ}^{({\rm av})} (\beta, \hat{m},r) = 
\frac{1}{3} Z_{QQ}^{(\overline{3})} (\beta, \hat{m},r) +
\frac{2}{3} Z_{QQ}^{(6)} (\beta, \hat{m},r)  \; .
\label{sum}
\end{equation}

\begin{figure}[t]
\epsfig{file=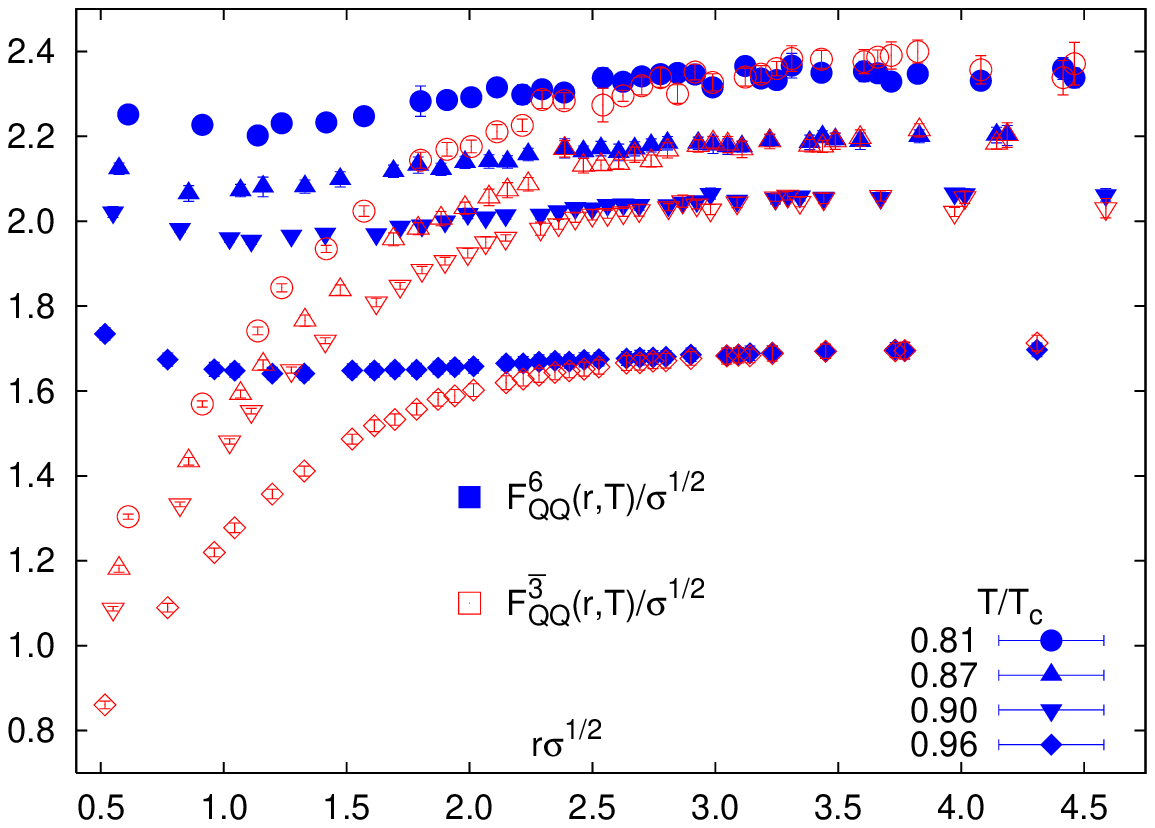,width=7.5cm}
\epsfig{file=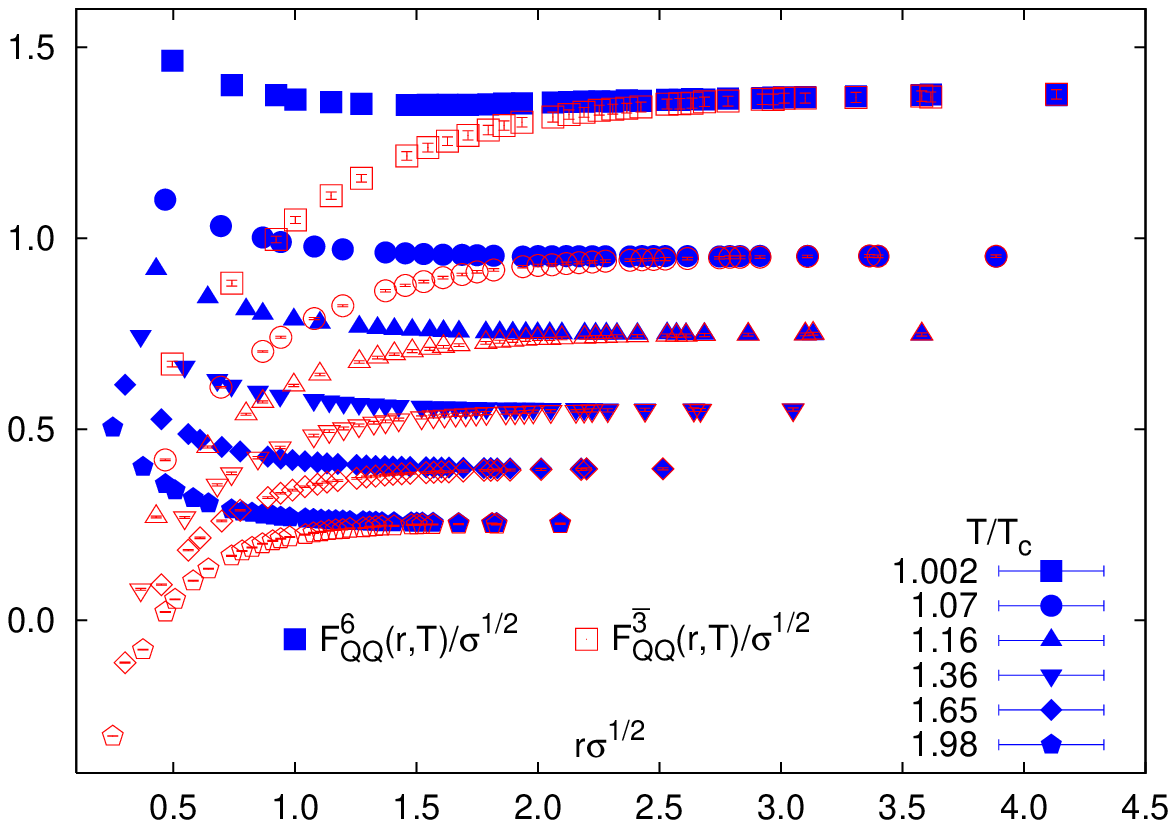,width=7.5cm}
\caption{Static diquark free energy in the anti-triplet
(open symbols)  and sextet  (filled symbols) channels
below (upper figure) and above (lower figure) the transition temperature
of 2-flavor QCD.}
\label{fig:36}
\end{figure}

It has already been observed in analogous studies of singlet and 
octet free energy in quark-antiquark systems that the large
distance behavior of free energies does not depend on the relative
color orientation of the two static sources. 
As shown in Fig.~\ref{fig:36} this is also reflected 
in the large distance behavior of diquark free energies.
The same gauge invariant asymptotic value is approached by $QQ$ free
energies calculated with sources in the anti-triplet and sextet 
representation. 

\subsection{Screening of anti-triplet diquarks}

We want to explore here in more detail the relation between the $r$-dependence 
of static diquark and quark-antiquark free energies. In leading order
perturbation theory the interaction in a $\bar{Q}Q$ singlet state
is twice as strong as in a $QQ$ anti-triplet state. This also carries over
to finite temperature perturbation theory where in leading order the Coulomb 
interaction is modified only through a Debye-screening term that reflects 
the in-medium properties of the exchanged gluons, 
\begin{eqnarray}
F^{(\overline{3})}_{QQ,pert}(r,T) &=& \frac{1}{2} F_{Q\bar Q,pert}^{(1)}(r,T)
\; +\; {\rm const.}
\nonumber \\
&\sim& -\frac{2}{3}\frac{\alpha(T)}{r}e^{-m_D(T) r} \; +\; {\rm const.}  
\end{eqnarray}
This suggest that at least at high temperatures the $r$-dependent 
$QQ$ and $\bar{Q}Q$ interaction part in the free energies obeys
Casimir scaling.  However, the gluonic screening cloud in a $QQ$ and
$\bar{Q}Q$ system will differ as a function of distance. While we just 
have deduced from the long distance behavior of free energies shown in
Fig.~\ref{fig:36} that at infinite distances the screening clouds contribute 
the same amount to the free energy of a diquark and quark-antiquark system,
respectively, 
this clearly has to be different at short distances. A $\bar{Q}Q$ system 
in a color singlet state does not require
a screening cloud at short distances; it is color neutral by
construction. However, an anti-triplet $QQ$ state needs to be 
screened even when the separation between both sources vanishes.
In fact, we expect that at short distances the free energy of the 
screening cloud for an anti-triplet $QQ$ state is identical to that
of a single static antiquark, $F_Q(T)$. 

\begin{figure}[t]
\epsfig{file=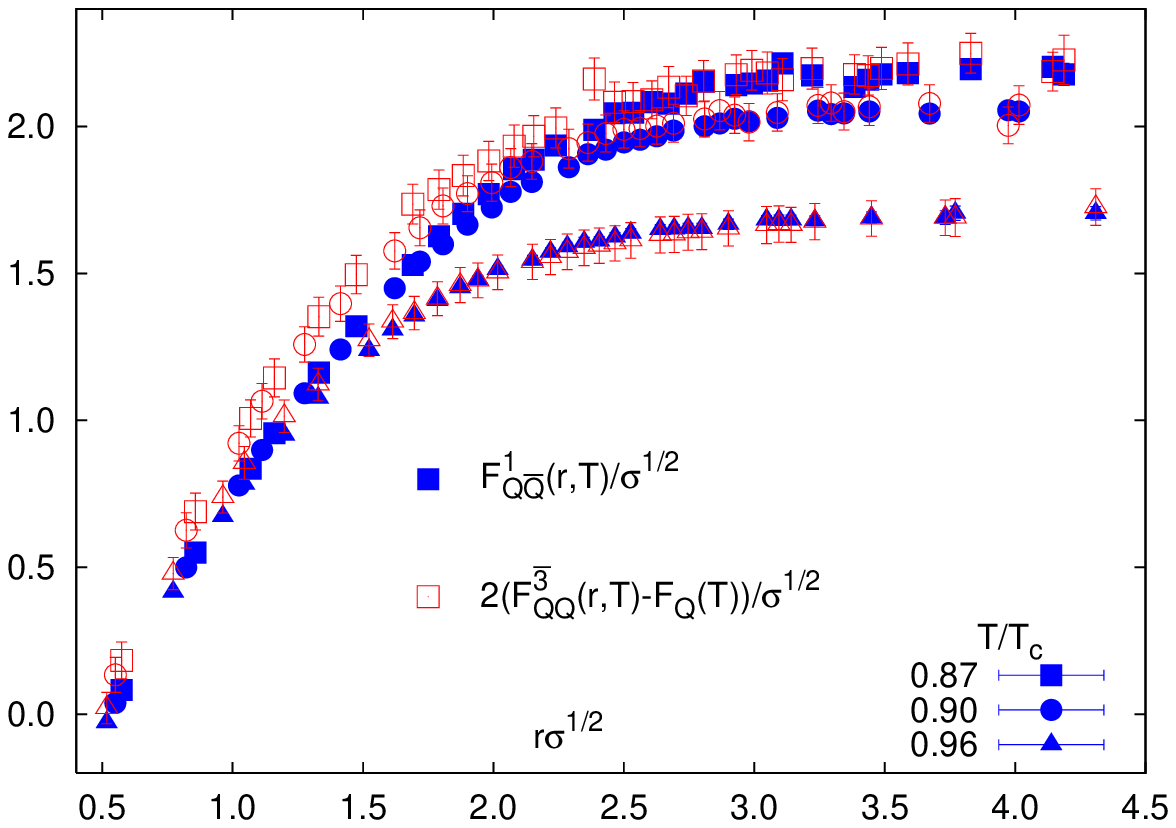,width=7.5cm}
\epsfig{file=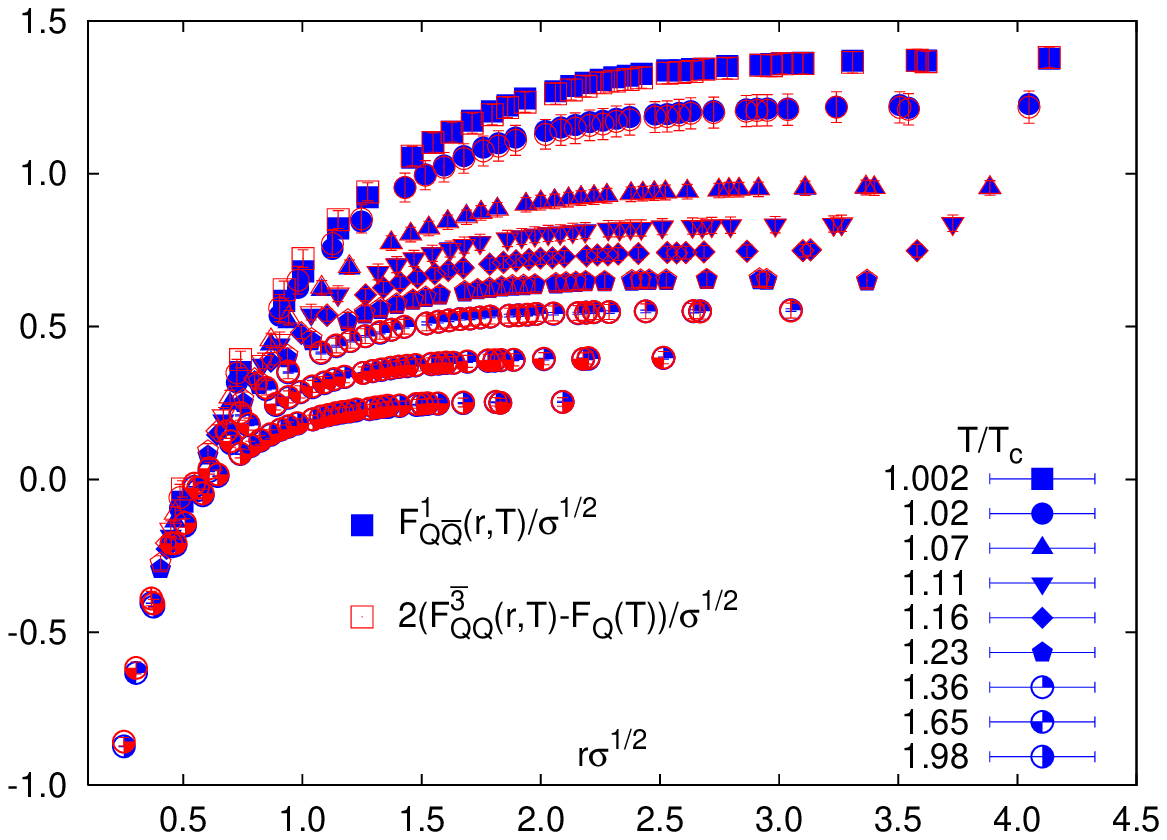,width=7.5cm}
\caption{Comparison of  anti-triplet diquark free energies (open symbols) 
and singlet free energies of a quark-antiquark system (filled symbols) 
below (upper figure) and above (lower figure) the transition temperature
of 2-flavor QCD.}
\label{fig:compare}
\end{figure}

The above considerations suggest a simple relation between free 
energies of anti-triplet $QQ$ states and color singlet $\bar{Q}Q$
systems, which should be valid at least at high temperature
as well as at large and short distances, respectively,
\begin{equation}
F_{Q\bar Q}^{(1)}(r,T) \simeq 2 (F^{(\overline{3})}_{QQ}(r,T)  - F_Q(T) ) \; .  
\label{QQdifference}
\end{equation}
In Fig.~\ref{fig:compare} we compare the left and right hand
sides of Eq.~\ref{QQdifference} at various temperatures. As can be seen
the equality indeed holds very well at all distances for temperatures above 
the transition
temperature, while some differences show up at intermediate distances
below $T_c$. Nonetheless, at short distances the equality in 
Eq.~\ref{QQdifference} holds at all temperatures 
which shows that indeed a well localized anti-triplet diquark system is 
screened in the same way as a single antiquark.
Moreover, we note here that Eq.~\ref{QQdifference} is found to hold also in 
the deconfined phase of SU(3) pure gauge theory \cite{Hubner}.

\subsection{Screening and string breaking}\label{sec_string_breaking}

We want to elaborate here a bit more in detail on the differences
seen in the screening of color singlet quark-antiquark and anti-triplet
diquark free energies below $T_c$. As noted before the contribution of
states with at least one additional 
quark is needed in the diquark partition function to project on the sector
of vanishing triality. At low temperature it is, in fact, a single quark
which dominates the additional contributions in a diquark partition
function for small separations of the two quarks. At large distances,
on the other hand, two antiquarks will be needed to screen the
color charges of the two well separated quarks. The situation in a quark
antiquark system is quite different. No additional quarks are needed for
screening of the sources at short distances, while a quark-antiquark 
pair is needed for screening at large distances, {\it i.e.} the 
net quark number will always vanish. 

At low temperature the screening mechanism at large distances
generally is referred to as string breaking. At high 
temperature such a screening mechanism based on the presence 
of dynamical quarks in the thermal medium or the vacuum 
becomes less relevant; screening with gluons becomes the dominant
mechanism. 
To explore the features of screening and string breaking and the change
of screening mechanisms at low and high temperature on a more 
quantitative level we have analyzed the net quark number induced in a 
thermal medium due
to the presence of a diquark system separated by a distance $r$. The
net quark number can be obtained from the static diquark free
energy, Eq.~\ref{eq:fx}, by considering its parametric dependence  
on the quark chemical potential through the fermion determinant as
indicated in Eq.~\ref{partition}. A derivative of 
$\ln Z_{QQ}^{(c)}(r,T)$ with
respect to $\mu/T$ evaluated at $\mu/T =0$ then yields the net quark
number,
\begin{eqnarray}
N_{QQ}^{(c)} (r,T) =
\left< N_q \right>_{QQ}
~&=&\frac{\left< N_q L^{(c)}_{QQ} (r,T) \right>}{\left< L^{(c)}_{QQ} (r,T)
\right>} \;,
\label{fn}
\end{eqnarray}
where $N_q$ is the quark number operator in 2-flavor QCD,
\be
N_q = \frac{1}{2}{\rm Tr}\left[ D^{-1} (\hat{m}, 0) 
\left( \frac{\partial D(\hat{m},\mu)}{\partial \mu} \right)_{\mu=0}
\right] \; .
\ee
Similarly we obtain the net quark number induced by a single static quark 
source,
\begin{eqnarray}
N_{Q} (T) =
\left< N_q \right>_{Q}
=\frac{\left< N_q  \tr P(\vec{0}) \right>}{\left< \tr P(\vec{0})
\right>} \;.
\label{fnq}
\end{eqnarray}

\begin{figure}[t]
\epsfig{file=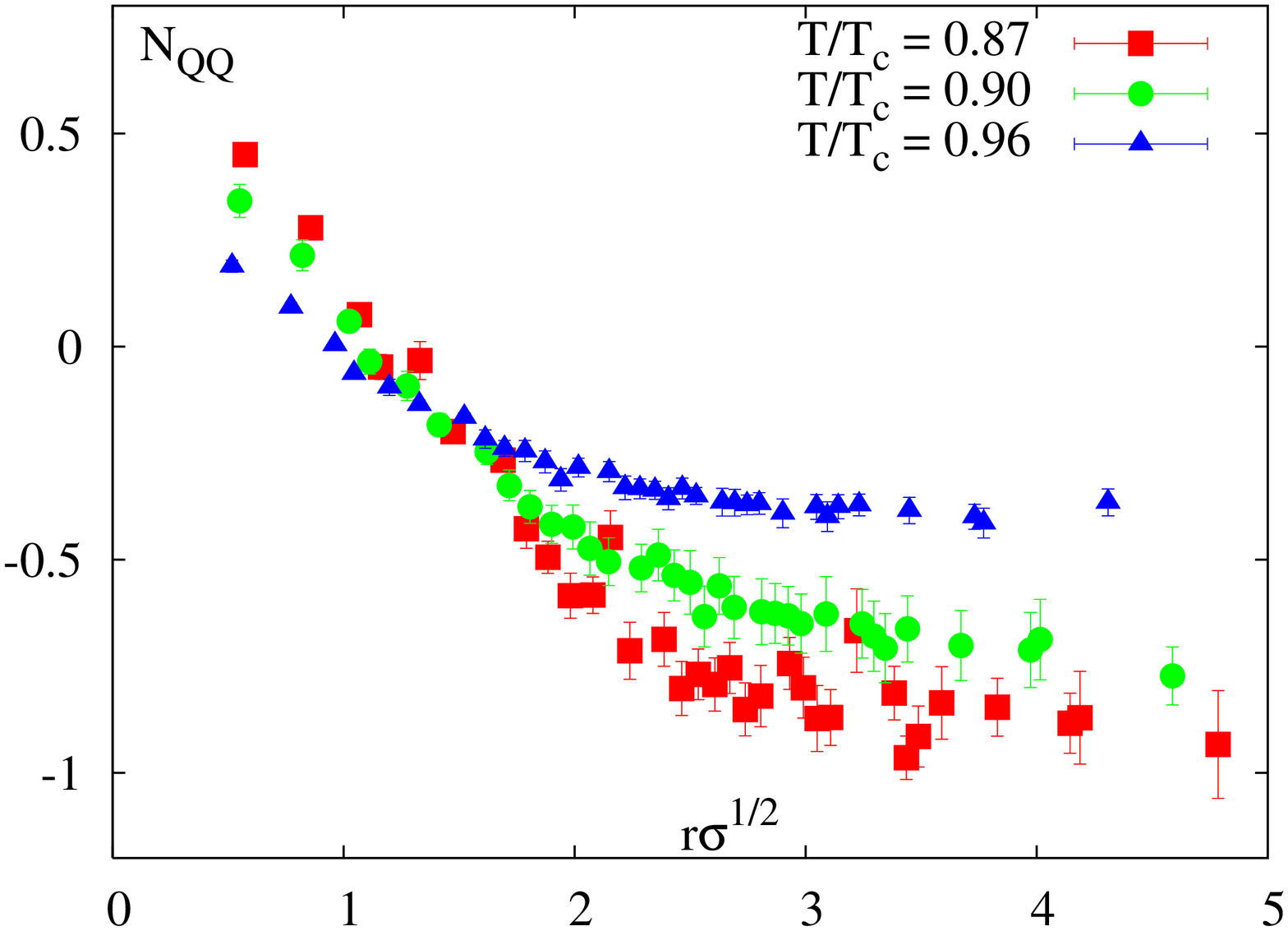,angle=270,width=7.5cm}
\epsfig{file=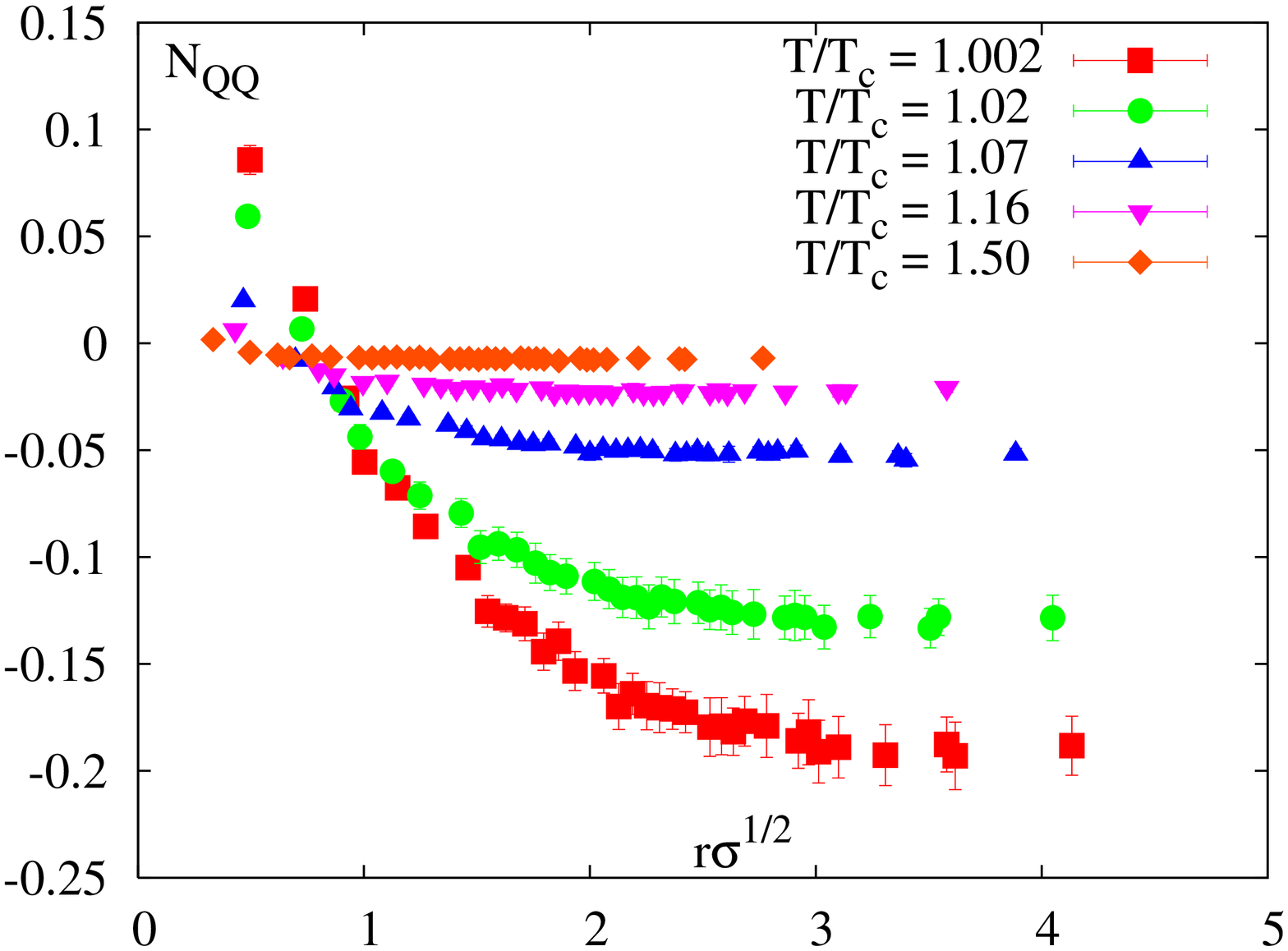,angle=270,width=7.5cm}
\caption{The net quark number induced by color averaged sources separated by a
distance $r\sqrt{\sigma}$ at temperatures below (upper plot) and 
above (lower plot) the transition temperature in 2-flavor QCD.}
\label{fig:plcqqav_c1}
\end{figure}

For a quark-antiquark system it is straightforward to see that the 
induced net quark number indeed is strictly zero for all temperatures,
{\it i.e.} $N_{\bar{Q}Q}^{(c)} (r,T) \equiv 0$. However, as argued above,
for a single quark or diquark system a net quark number has 
to be induced to generate a state with vanishing triality. In 
Fig.~\ref{fig:plcqqav_c1} we show
the net quark number induced by color averaged diquark sources separated
by a distance $r$. We observe a clear separation between the short and
long distance behavior of this observable. While for small separation
of the sources a net quark number ($N_{QQ} > 0$) is induced, a net
antiquark number ($N_{QQ} < 0$) is induced at large separations. This
effect becomes more pronounced with decreasing temperature. Above the
transition temperature, however, the net quark number rapidly goes to
zero. 

\begin{figure}[t]
\epsfig{file=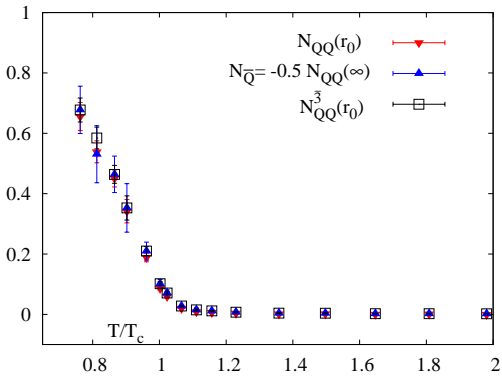,angle=270,width=7.5cm}
\caption{The net quark number induced by color averaged sources separated a
distance $r_0a=1/4$ (lower triangles) and a static antiquark sources,
$N_{\bar{Q}}$ (upper triangle), which equals minus half the net quark 
number induced by a diquark system at infinite separation ($N_{QQ}(\infty)$).
Also shown is the net quark number induced by a anti-triplet diquark system
separated a distance $r_0a=1/4$ (squares). 
}
\label{fig:Nq_asymptotic}
\end{figure}

The properties of $N_{QQ}$ become more apparent when we consider
also the net quark numbers induced by static sources in an anti-triplet
state. In Fig.~\ref{fig:Nq_asymptotic} we show the values for the induced
net quark number in color averaged and anti-triplet diquark systems in the large
distance limit as well as for the shortest distance available, $rT=1/N_\tau$. 
At all temperatures we find that the net quark number induced for small
separations of the sources in the anti-triplet channel is positive and
equals the
net quark number needed to 'neutralize' a static antiquark source.
At large separations, on the other hand, it is minus twice that 
needed to 'neutralize' static antiquark source.  At short distances a $QQ$ 
anti-triplet diquark thus is 'neutralized' by a light 
quark from the thermal medium, while at
large distances both quark sources are individually 'neutralized' by a
light antiquark. While at finite temperature $N_{QQ}$ also receives
contributions from other (multi-quark) states, this mechanism becomes
increasingly dominant at low temperature. In fact, the data shown in
Figs.~\ref{fig:plcqqav_c1} and \ref{fig:Nq_asymptotic} support the expected
limiting behavior,
\begin{eqnarray}
\lim_{T\rightarrow 0} N_{QQ}(r,T) &=& \begin{cases}
~1 &, \; r< r_c    \\
-2 &, \; r> r_c \\
\end{cases}    \; ,
\label{limits}
\end{eqnarray}
with $r_c$ being a typical hadronic scale, $r_c\simeq 1.5/\sqrt{\sigma} \simeq
0.8~$fm, that characterizes the string breaking radius at zero temperature.

In all cases considered here the induced net quark number rapidly becomes 
small at
high temperature. In fact, already for $T\gsim 1.2$ the deviations from zero
are smaller than 0.02 at all distances. Thermal fluctuations of the quark
number and the largely available number of thermal gluons thus seem to
be sufficient to screen the static color sources at high temperature.

\subsection{Screening of sextet diquarks}
\label{subsec:sextet_screening}

\begin{figure}[t]
\epsfig{file=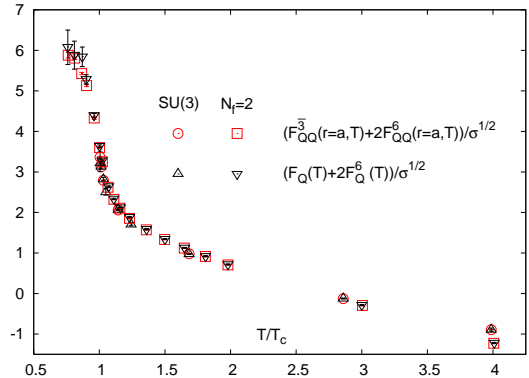,width=7.5cm}
\caption{Comparison of  anti-triplet and sextet diquark free energies at
short distances with the free energy of static fermions in triplet
and sextet representations. The particular combination shown eliminates
the short distance Coulomb terms contributing to the diquark free 
energies. The sextet free energies of static quarks have been computed in 
Ref.~\cite{huebner}.}
\label{fig:f63}
\end{figure}

The considerations of the previous section can also be carried over to 
sextet diquark systems. The perturbative one-gluon exchange among
quarks in a sextet representation is repulsive. Also this interaction
is screened at finite temperature. At high temperature we thus expect
to find for the genuine 'interaction part' in the sextet free energy, 
\begin{equation}
F^{(6)}_{QQ}(r,T) \sim -\frac{1}{4} F_{Q\bar Q}^{(1)}(r,T)
\sim \frac{1}{3}\frac{\alpha(T)}{r}e^{-m_D(T) r} \; .
\label{QQ6}
\end{equation}
While the large distance behavior of the sextet and anti-triplet
free energies is identical (see Fig.~\ref{fig:36}), we again 
expect that at short distances the sextet diquark state
is screened differently from the anti-triplet. In fact, we expect that
screening of a sextet diquark at short distances is identical to 
screening of a single fermion in a color sextet representation,
{\it i.e.} when combining sextet and anti-triplet free energies in
proportion to  the corresponding Casimir factors such that
the repulsive and attractive Coulombic contributions cancel we expect 
to find 
\begin{equation}
\lim_{r\rightarrow 0} (2 F^{(6)}_{QQ}(r,T) + F^{(\overline{3})}_{QQ}(r,T))
=  2 F^{(6)}_Q (T) + F_Q(T)
\label{F63}
\end{equation} 
where $F^{(6)}_Q (T)$ is the free energy of a fermion in a color 
sextet representation given by the corresponding Polyakov loop
expectation value,
\begin{equation}
{\rm e}^{-F^{(6)}_Q /T} = \left\langle \frac{1}{N_\sigma^3}\sum_{\vec{x}}L_6(\vec{x})\right\rangle,
\label{F6}
\end{equation} 
where 
\begin{equation}
L_6(\vec{x}) = \frac{3}{2}\left(\tr\,P(\vec{x})\right)^2-\frac{1}{2}\tr\,P^{\dagger}(\vec{x})
\label{F6_aux}
\end{equation} 
is obtained from standard group theoretical relations.
Eq.~\ref{F63} is indeed fulfilled at all temperatures in 2-flavor QCD as
well as  in the deconfinement phase in SU(3) pure gauge theory, as can be seen
in Fig.~\ref{fig:f63}. We note that the expectation value of a Polyakov loop
in the sextet representation is, of course, gauge invariant. The results
on the short distance behavior of sextet as well as anti-triplet quark-quark 
free energies thus reflect gauge invariant screening properties of 
colored diquark systems.   

\section{Heavy Diquark bound states in the Quark Gluon Plasma?}

The analysis of diquark and quark-antiquark free energies presented in the 
previous section suggests that aside from the different screening mechanisms 
for $\bar{Q}Q$ and $QQ$ systems the r-dependence of the free energies obeys
Casimir scaling above $T_c$; {\it i.e.} the interaction in a diquark system
is only half as strong as in a $\bar{Q}Q$ system. An immediate consequence
of this is that the temperature dependent coupling defined in terms
of $QQ$ systems,

\begin{equation}
\alpha_{\rm eff}   
= \frac{3 r^2}{2} \frac{{\rm d} F_{QQ}^{(\bar{3})}(r,T)}{{\rm d}r} \; ,
\end{equation}
coincides with that extracted previously from color singlet free
energies \cite{Zan2}. 

The close relation between color singlet quark-antiquark free energies 
and anti-triplet diquark free energies, Eq.~\ref{QQdifference},
of course also carries over to the corresponding energies ($U$) and 
entropies ($S$) 
deduced from the free energy, $F=U-TS$, through appropriate partial
derivatives with respect to temperature at fixed separation, $r$,
of the sources.
Free energies as well as combinations of energy and entropy of static quark 
systems have been used to construct 
temperature dependent, effective potentials for heavy quark systems
that allow to model the temperature dependence of heavy quark bound
states in the high temperature phase of QCD. Although such constructions
are not rigorous and will include model assumptions, it seems to be 
clear from the analysis presented in the previous section that any
potential constructed for color singlet quark-antiquark systems also puts
stringent constraints on the interaction in a diquark system. We expect
to have
\begin{equation} 
V_{QQ} (r,T) = \frac{1}{2} V_{\bar{Q}Q} (r,T) \; .
\label{QQpotential}
\end{equation}
If a tightly bound $\bar{Q}Q$ systems can exist at high temperature
it thus is conceivable that also $QQ$ states may persist in some 
temperature regime above $T_c$.
As we have seen in the previous section the color charge of such 
a $QQ$ system will in addition be screened by a gluon cloud.

Using Eq.~\ref{QQpotential} we have solved the Schr\"odinger equation
for diquark states formed from charm and bottom quarks respectively,
\begin{equation}
\left[ 2m_a + \frac{1}{m_a}\nabla^2 + V_{x}(r,T) \right]
\Phi_i^a = M_i^a(T)\;
\Phi_i^a\; , \; a\; =\; {\rm c,}\; {\rm b}.
\end{equation}
As we do not want to go into a detailed modeling of heavy quark bound 
states but rather want to discuss some generic features expected to
hold for $QQ$-systems given some knowledge on the behavior of 
$\bar{Q}Q$-systems,
we have used a generic ansatz for the heavy quark potential, which
has been used 
previously also in studies of charmonium and bottomonium systems,
\begin{equation}
V_{\bar{Q}Q}(r,T) = \frac{\sigma}{\mu} \left( 1-{\rm e}^{-\mu(T)r} \right) -
\frac{\alpha}{r} {\rm e}^{-\mu(T)r} \; . \;
\label{pot}
\end{equation}
Here we have used parameters for the coupling in the Coulomb term and
the linear rising confinement term of the zero temperature  
potential that describe well the lattice results for the static quark
potential, $(\alpha, \sigma)=(0.385, 0.224{\rm GeV^{-2})}$ and 
adjusted the screening mass, $\mu(T)$, 
such that the finite temperature potential reproduces the melting 
temperature of the charmonium ground
state, $J/\psi$, at a temperature consistent with lattice studies of spectral
functions, $T_{\rm dis}(\bar{c}c)\simeq 1.5 T_c$ 
\cite{Hatsuda,ourmem,Jakovac:2006sf}. 
With this ansatz we find
dissociation temperatures for ground states in the bottomonium system 
($\Upsilon$) and the corresponding diquark states given in 
Table~\ref{tab:boundstates}. As can be seen this analysis suggests that 
$cc$ states are unlikely to exist in the high temperature phase of QCD.
They dissolve already at $T_c$. However, $bb$ states dissolve only 
at temperatures of about $2 T_c$. 

\begin{table}[t]
\begin{center}
\vspace{0.5cm}
\begin{tabular}{|c||c|c||c|c|}
\hline
state & $\bar{c}c\; (J/\psi)$ & $cc$ & $\bar{b}b\; (\Upsilon)$ & $bb$ \\
\hline
$E_{\rm dis}$~[GeV] & 0.06 & 0& 0.3& 0.07 \\
$T_{\rm dis}/T_c$ & 1.5 & 1.0& 3.2 & 2.1 \\
\hline
\end{tabular}
\end{center}
\caption{Dissociation temperatures, $T_{\rm dis}$, for heavy quark-antiquark
and diquark bound states in units of the transition temperature $T_c$ and
dissociation energies for these states at $T_c$.}
\label{tab:boundstates}
\end{table}

\section{Conclusions}

We have presented results on free energies of static diquark systems.
A detailed comparison with corresponding results for free energies
of static quark-antiquark systems shows that the relative magnitude
of the r-dependent part of these free energies obeys Casimir scaling
although at short distances the screening of diquark states is quite 
different from that of quark-antiquark states; colored diquark systems
at short distances are screened like static quark sources in the same
color representation.  Using these results
in a model for heavy diquark bound states we showed that it is likely
that in a quark gluon plasma $bb$ states can form at temperatures
well above the transition temperature while $cc$ states are unlikely to
exist above $T_c$. 

\section*{Acknowledgments}
This work has been supported in part by contract DE-AC02-98CH1-886
with the U.S. Department of Energy, the Helmholtz Gesellschaft under grant
VI-VH-041 and the Deutsche Forschungsgemeinschaft under grant GRK 881.


\end{document}